\documentclass[11pt,a4paper]{article}

\usepackage{epsfig,amsmath,amssymb,cite}

\begin{document}

\title{Thin--shell wormholes with a generalized Chaplygin gas in Einstein--Born--Infeld theory} 
\author{Ernesto F. Eiroa$^{1, 2}$\thanks{e-mail: eiroa@iafe.uba.ar}, Griselda Figueroa Aguirre$^{1}$\thanks{e-mail: gfigueroa@iafe.uba.ar}\\
{\small $^1$ Instituto de Astronom\'{\i}a y F\'{\i}sica del Espacio, C.C. 67, 
Suc. 28, 1428, Buenos Aires, Argentina}\\
{\small $^2$ Departamento de F\'{\i}sica, Facultad de Ciencias Exactas y 
Naturales,} \\ 
{\small Universidad de Buenos Aires, Ciudad Universitaria Pab. I, 1428, 
Buenos Aires, Argentina}} 

\maketitle

\begin{abstract}
We construct spherically symmetric thin--shell wormholes supported by a generalized Chaplygin gas in Born--Infeld electrodynamics coupled to Einstein gravity, and we analyze their stability under radial perturbations. For different values of the Born--Infeld parameter and the charge, we compare the results with those obtained in a previous work for Maxwell electrodynamics. The stability region in the parameter space reduces and then disappears as the value of the Born--Infeld parameter is modified in the sense of a larger departure from Maxwell theory. \\

\noindent 
PACS number(s): 04.20.Gz, 04.40.Nr, 98.80.Jk

\end{abstract}

\section{Introduction}\label{intro} 

Traversable Lorentzian wormholes are theoretical objects which have received great attention in the last two decades. These objects have a throat that connects two regions of the same universe or two different universes \cite{motho,visser} and, in general relativity, they are characterized by being threaded by matter that violates the null energy condition \cite{motho,visser,hovis} in order to allow material systems travel through them. The amount of this exotic matter can be made arbitrary small \cite{viskardad}, but at the expense of large pressures at the throat \cite{lst}. Traversable wormholes can be constructed \cite{visser} by using the well known  thin--shell formalism  -commonly employed to model layers in different physical contexts, including modern cosmology (e.g. braneworlds) and gravastars-, consisting in the cut and paste of two manifolds to form a new one, with a shell at the joining surface corresponding the throat, where the fulfillment of the flare--out condition is required. These 
wormholes are of particular interest for their simplicity, which facilitates the stability analysis, and because the presence of exotic matter is confined to the shell. For these reasons, they are widely studied in the literature. Models of wormholes with a continuous energy-stress tensor at the throat usually also require a cut and paste procedure to confine the exotic matter and/or obtain a suitable asymptotic behavior. Stability studies of spherically symmetric thin--shell wormholes, with a linearized equation of state at the throat, have been performed under radial perturbations(\cite{poisson,ishak-eirom-lobo,eir,dilem,lmv} and references therein). Plane and cylindrical thin--shell wormholes were also considered in recent years (see, for example, Refs. \cite{plane,dilem,cil}).

Born--Infeld electrodynamics \cite{borninf} is a non--linear theory proposed in order to avoid the infinite self energies of charged point particles arising in Maxwell theory, and it is the only non--linear theory without birefringence. The spherically symmetric solution in general relativity coupled to Born--Infeld electrodynamics was obtained by Hoffmann \cite{hoffmann}; this solution failed to be a suitable model for the electron, corresponding instead to  a black hole. Born--Infeld type actions have appeared in  low energy string theory \cite{nle}, leading to an increase in the interest of non--linear electrodynamics. Maxwell and Born--Infeld theories have electric-magnetic duality invariance \cite{gibbons}, property not shared with other electromagnetic theories. The geodesic structure of Einstein--Born--Infeld black holes was studied in Ref. \cite{nora}. 
The linearized stability of spherical shells and thin--shell wormholes under radial perturbations was recently considered within this theory \cite{shellbi}.

In the framework of general relativity, the accelerated expansion of the Universe violates the strong energy condition. Several models of exotic matter, proposed in cosmology \cite{matt}, have been also adopted in wormhole spacetimes. One of them, the Chaplygin gas \cite{aero,chap} was used as the exotic matter supporting wormholes \cite{lobo73,chaply07,chaplywh,chaply09,chaply11}. In particular, a generalized Chaplygin gas was taken as the exotic matter at the throat of thin--shell wormholes in Refs. \cite{chaply09,chaply11}.

In the present work, we construct thin--shell wormholes with a generalized Chaplygin gas in Einstein--Born--Infeld theory, and we analyze their stability under perturbations that conserve the symmetry. As in Ref. \cite{chaply09}, we adopt the generalized Chaplygin gas equation of state at the throat mainly because of its interest in modern cosmology. This fluid introduces two parameters and non trivial complications in the equations. Here we extend to Born-Infeld electrodynamics the analysis performed previously using Maxwell theory \cite{chaply09}. We compare the new results with those obtained in Ref. \cite{chaply09}. The main outcome of our paper is that the stability region reduces in size and finally fades away as the Born-Infeld parameter takes values in the direction of a larger deviation from Maxwell electrodynamics. 
The paper is organized as follows: in Sec.  \ref{tswh}, the wormhole construction is done; in Sec. \ref{stab}, the stability of static configurations is analyzed; finally, in Sec. \ref{conclu} a summary is presented.

\section{Wormhole construction: general equations}\label{tswh}

The action of Born--Infeld electrodynamics coupled to Einstein gravity is given by (in units such as $G=c=1$)
\begin{equation}
S=\int d^4x \sqrt{g}\left( \frac{1}{16\pi } R + L \right) ,
\label{bi1} 
\end{equation} 
where $R$ is the scalar of curvature, $g=det|g_{\mu \nu}|$, and $L$ depends on the electromagnetic tensor in a non--linear form:
\begin{equation}
L=\frac{1}{4\pi b^{2}}\left( 1-\sqrt{1+
\frac{1}{2}F_{\sigma \nu }F^{\sigma \nu }b^{2}-
\frac{1}{4}\mbox{}^{*}F_{\sigma \nu }F^{\sigma \nu }b^{4}}\right) ,
\label{bi0b}
\end{equation}
with $F_{\sigma \nu }=\partial _{\sigma }A_{\nu } -\partial _{\nu }A_{\sigma }$ the electromagnetic tensor, $\mbox{}^{*}F_{\sigma \nu } =\tfrac{1}{2}\sqrt{-g}\, \varepsilon _{\alpha \beta \sigma \nu }F^{\alpha \beta }$ the Hodge dual of $F_{\sigma \nu }$, and $\varepsilon _{\alpha \beta \sigma \nu }$ the Levi--Civita symbol. The parameter $b$ indicates how much Born--Infeld and Maxwell electrodynamics differ; $b^{-1}$ is the maximum of the electric field. In the limit $b\rightarrow 0 $, the Maxwell lagrangian is recovered.

The field equations, obtained from the action (\ref{bi1}), have the vacuum spherically symmetric solution \cite{gibbons,nora}: 
\begin{equation} 
ds^2=-\psi (r) dt^2+\psi (r)^{-1} dr^2+r^2(d\theta^2 + \sin^2\theta d\phi^2),
\label{metricaBI}
\end{equation}
where $r>0$ is the radial coordinate, $0\le \theta \le \pi$ and $0\le \varphi<2\pi $ are the angular coordinates, and $\psi$ has the form:
\begin{equation} 
\psi (r) = 1 - \frac{2M}{r} + \frac{2}{3b^2}\left\lbrace r^2 - \sqrt{r^4 + b^2 Q^2} + \frac{\sqrt{|bQ|^3}}{r} F \left[ \arccos \left( \frac{r^2 - |bQ|}{r^2+|bQ|} \right) ,\frac{\sqrt{2}}{2} \right] \right\rbrace ,
\label{psiBI}
\end{equation}
with $M$ the mass, $Q$ the charge, and $F(\gamma ,k)$ the elliptic integral of the first kind: $F(\gamma , k) = \int _{0}^{\sin \gamma }[(1-z^{2})(1-k^{2}z^{2})]^{-1/2}dz =\int _{0}^{\gamma }(1-k^{2}\sin ^{2}\phi )^{-1/2}d\phi$. The geometry is singular at $r=0$; the position of the horizons, determined by the zeros of $\psi (r)$, have to be calculated numerically. The Schwarzschild metric is recovered if $Q=0$ and the Reissner--Nordstr\"{o}m geometry is obtained by taking the limit $b\rightarrow 0 $ (for more details, see Ref. \cite{nora}).

We start from the metric showed in Eq. (\ref{metricaBI}) to construct thin--shell wormholes by using the  Darmois--Israel formalism \cite{daris}. We need to take a radius $a$ larger than the event horizon $r_h$ in order to avoid the presence of the singularity and the horizons. We cut two identical copies of the region with $r\geq a$:
\begin{equation} 
\mathcal{M}^{\pm }=\{X^{\alpha }=(t,r,\theta,\varphi)/r\geq a\},  \label{e2}
\end{equation}
and paste them at the hypersurface
\begin{equation} 
\Sigma \equiv \Sigma ^{\pm }=\{X/F(r)=r-a=0\},  \label{e3}
\end{equation}
to create a new geodesically complete manifold $\mathcal{M}=\mathcal{M}^{+} \cup \mathcal{M}^{-}$. If this construction satisfies the flare-out condition, the manifold represents a wormhole with two regions connected by a throat of radius $a$, which corresponds to the surface of minimal area. The flare--out condition is satisfied in our case, because $\psi '(a)=2a>0$. A global radial coordinate can be defined on $\mathcal{M}$ by using the proper radial distance: $l=\pm \int_{a}^{r}\sqrt{1/\psi (r)}dr$, the signs $\pm $ correspond respectively, to $\mathcal{M}^{+}$ and $\mathcal{M}^{-}$, and the throat is located in $l=0$. At the throat $\Sigma $ we can define the coordinates $\xi ^{i}=(\tau ,\theta,\varphi )$, with $\tau $ the proper time on the shell. The throat radius is a function of time: $a(\tau)$. A Birkhoff theorem holds for the metric adopted in the construction \cite{gibbons}, so the geometry remains static outside the throat and no gravitational waves are present. Adopting the orthonormal basis $\{ 
e_{\hat{\tau}}=e_{\tau }, e_{\hat{\theta}}=a^{-1}e_{\theta }, e_{\hat{\varphi}}=(a\sin \theta )^{-1} e_{\varphi }\} $, it is easy to obtain the second fundamental forms (or extrinsic curvature) associated with the two sides of the shell:
\begin{equation} 
K_{\hat{\theta}\hat{\theta}}^{\pm }=K_{\hat{\varphi}\hat{\varphi}}^{\pm
}=\pm \frac{1}{a}\sqrt{\psi (a) +\dot{a}^2},
\label{e6}
\end{equation}
and
\begin{equation} 
K_{\hat{\tau}\hat{\tau}}^{\pm }=\mp \frac{\psi '(a)+2\ddot{a}}{2\sqrt{\psi(a)+\dot{a}^2}},
\label{e7}
\end{equation}
where a prime represents a derivative with respect to $r$ and the dot with respect to $\tau$. With the following definitions:
$[K_{_{\hat{\imath}\hat{\jmath}}}]\equiv K_{_{\hat{\imath}\hat{\jmath}}}^{+}-K_{_{\hat{\imath}\hat{\jmath}}}^{-}$, $K=tr[K_{\hat{\imath}\hat{\jmath }}]=[K_{\; \hat{\imath}}^{\hat{\imath}}]$ 
and with the surface stress-energy tensor
$S_{_{\hat{\imath}\hat{\jmath} }}={\rm diag}(\sigma ,p_{\hat{\theta}},p_{\hat{\varphi}})$, where $\sigma$ is the surface energy density and $p_{\hat{\theta}}$, $p_{\hat{\varphi}}$ are the transverse pressures, the Einstein equations on the shell can be reduced to Lanczos equations:
\begin{equation} 
-[K_{\hat{\imath}\hat{\jmath}}]+Kg_{\hat{\imath}\hat{\jmath}}=8\pi 
S_{\hat{\imath}\hat{\jmath}};
\label{e8}
\end{equation}
then we have that
\begin{equation} 
\sigma=-\frac{\sqrt{\psi(a)+\dot{a}^2}}{2\pi a},
\label{e9}
\end{equation}
and
\begin{equation}
p=p_{\hat{\theta}}=p_{\hat{\varphi}}=\frac{\sqrt{\psi(a)+\dot{a}^2}}{8\pi} \left[\frac{2}{a} + \frac{2\ddot{a}+\psi '(a)}{\psi(a)+\dot{a}^2}\right] .
\label{e10}
\end{equation}
It can be seen from Eq. (\ref{e9}) that  $\sigma <0$, which indicates the presence of exotic matter. Here we model this exotic matter with a generalized Chaplygin gas on the shell $\Sigma $. In this gas, the pressure has opposite sign to the energy density. Then, the equation of state at the throat can be written in the following form: 
\begin{equation}
p=\frac{A}{|\sigma |^{\alpha}},
\label{e11} 
\end{equation} 
where $A>0$ and $0<\alpha\le 1$ are constants. When $\alpha=1$ the ordinary Chaplygin gas equation of state $p=-A/\sigma$ is recovered. The dynamical evolution of the wormhole throat can be obtained by replacing Eqs. (\ref{e9}) and (\ref{e10}) into Eq. (\ref{e11}):
\begin{equation}
\left\{ [2\ddot{a}+\psi' (a)]a^2+[\psi(a)+\dot{a}^2]2a \right\}[2a]^{\alpha}-2A[4\pi a^{2}]^{\alpha +1}[\psi (a)+\dot{a}^2]^{(1-\alpha)/2}=0.
\label{e12} 
\end{equation}
In this way, we have found a differential equation that should be satisfied by the radius of throat for thin--shell wormholes in Einstein--Born--Infeld theory, threaded by exotic matter with the equation of state of a generalized Chaplygin gas.

\section{Stability of static solutions}\label{stab}

In the case of static wormholes, from Eqs. (\ref{e9}) and (\ref{e10}), the surface energy density and pressure are given by
\begin{equation}
\sigma_{0}=-\frac{\sqrt{\psi(a_{0})}}{2\pi a_{0}},
\label{e13}
\end{equation}
and
\begin{equation}
p_{0}=\frac{\sqrt{\psi(a_{0})}}{8\pi}\left[\frac{2}{a_{0}} + \frac{\psi '(a_{0})}{\psi(a_{0})}\right] .
\label{e14}
\end{equation}
If static solutions exist for a given set of parameters, they should satisfy Eq. (\ref{e12}) evaluated in $a_{0}$:
\begin{equation}
(2a_0)^\alpha [\psi '(a_0)a_0^2+2\psi(a_0)a_0]-2A[4\pi a_0^2]^{\alpha +1} \psi(a_0)^{(1-\alpha)/2}=0.
\label{e12eval}
\end{equation}
From Eqs. (\ref{e9}) and (\ref{e10}) it is easy to verify the conservation equation:
\begin{equation}
\frac{d}{d\tau }\left( \sigma \mathcal{A}\right) +p\frac{d\mathcal{A}}{d\tau }=0,
\label{p1}
\end{equation}
where $\mathcal{A}=4\pi a^2$ is the area of the wormhole throat. In Eq. (\ref{p1}), the first term represents the internal energy change of the throat and the second the work done by the internal forces of the throat. We can rewrite Eq. (\ref{p1}) in the form:
\begin{equation}
\dot{\sigma}=-2\left( \sigma + p\right) \frac{\dot{a}}{a},
\label{p2}
\end{equation}
which can be integrated to give
\begin{equation}
\ln \frac{a}{a(\tau_{0})}=-\frac{1}{2} \int_{\sigma(\tau_{0})}^{\sigma} \frac{d\sigma}{\sigma+p(\sigma)}.
\label{p3}
\end{equation}
This equation can be formally inverted to obtain $\sigma = \sigma (a)$. Then we replace $\sigma(a) $ in Eq. (\ref{e10}) to find  the equation that determines completely the dynamics of the throat:
\begin{equation}
\dot{a}^{2}=-V(a)=-\left\{ \psi(a)-\left[2\pi a\sigma (a)\right] ^{2}\right\},
\label{p4}
\end{equation}
where $V(a)$ can be interpreted as a potential, which can be expanded in a second order Taylor series around the radius $a_0$ of the static solution, in order to analyze the stability:
\begin{equation}
V(a)=V(a_{0})+V'(a_{0})(a-a_{0})+\frac{V''(a_{0})}{2}(a-a_{0})^{2}+O(a-a_{0})^{3}.
\label{p6}
\end{equation}
The first and second derivatives of $V(a)$ are given by
\begin{equation}
V'(a)= \psi '(a)+8a\pi ^{2}\sigma (a)\left[ \sigma (a)+ 2 p(a) \right],
\label{p7}
\end{equation}
\begin{equation}
V''(a)=\psi ''(a)-8\pi ^2\left\lbrace \left[ \sigma (a)+2p(a)\right] ^{2} +2\sigma (a) \left[ \sigma (a)+p(a)\right] \left[ 1+2\frac{\alpha p(a)}{|\sigma (a)|}\right] \right\rbrace,
\label{p8}
\end{equation}
where we have used that $a\sigma '=-2(\sigma +p)$. By replacing Eqs. (\ref{e13}) and (\ref{e14}) in Eqs. (\ref{p4}), (\ref{p7}), and (\ref{p8}), we have that $V(a_{0})=V'(a_{0})=0$, and 
\begin{equation}
V''(a_{0})=\psi ''(a_{0})+\frac{(\alpha -1) [\psi '(a_{0})]^2 }{2\psi (a_{0})} + \frac{\psi '(a_{0})}{a_{0}}- \frac{2(\alpha +1)\psi (a_{0})}{a_{0}^2}.
\label{p9}
\end{equation}
From this last equation we obtain the stability condition for perturbations conserving the spherical symmetry of the geometry: the wormhole is stable if and only if $V''(a_{0})>0$.

By using Eq. (\ref{e12eval}), we can find the possible throat radii $a_{0}$, for different values ​​of the Born--Infeld parameter $b$, the constant  $A$, the exponent $\alpha$, the mass $M$ and  the charge $Q$. Since the inequality $V''(a_0)>0$, which determines whether the solution with radius $a_0$ is stable, is very complicated from an algebraic point of view, the results (obtained numerically) are presented graphically in Figs. 1-6, in which standard software was used and we have chosen the most representative figures. The stable solutions are shown with solid lines, while the dotted lines correspond to unstable configurations. The regions that have no physical meaning are shaded in gray. The results present an important change around $Q_c/M$, where $Q_{c}$ is the critical charge, corresponding to the value of charge from which the original metric used in the construction  has no horizons. The quotient $Q_c/M$ only depends on the parameter $b/M$. For a fixed value of $b/M$, the event horizon has a radius 
which decreases as $|Q|/M$ grows, and it fades out for values of $|Q|/M$ larger than $Q_c/M$. The values of $Q_c/M$ are those for which $\psi (r_h)=0$ and $\psi' (r_h)=0$ (i.e., double root of $\psi (r)$), to be obtained numerically.

\begin{figure}[t!]
\centering
\begin{minipage}{0.475\textwidth}
\centering
\includegraphics[width=\textwidth]{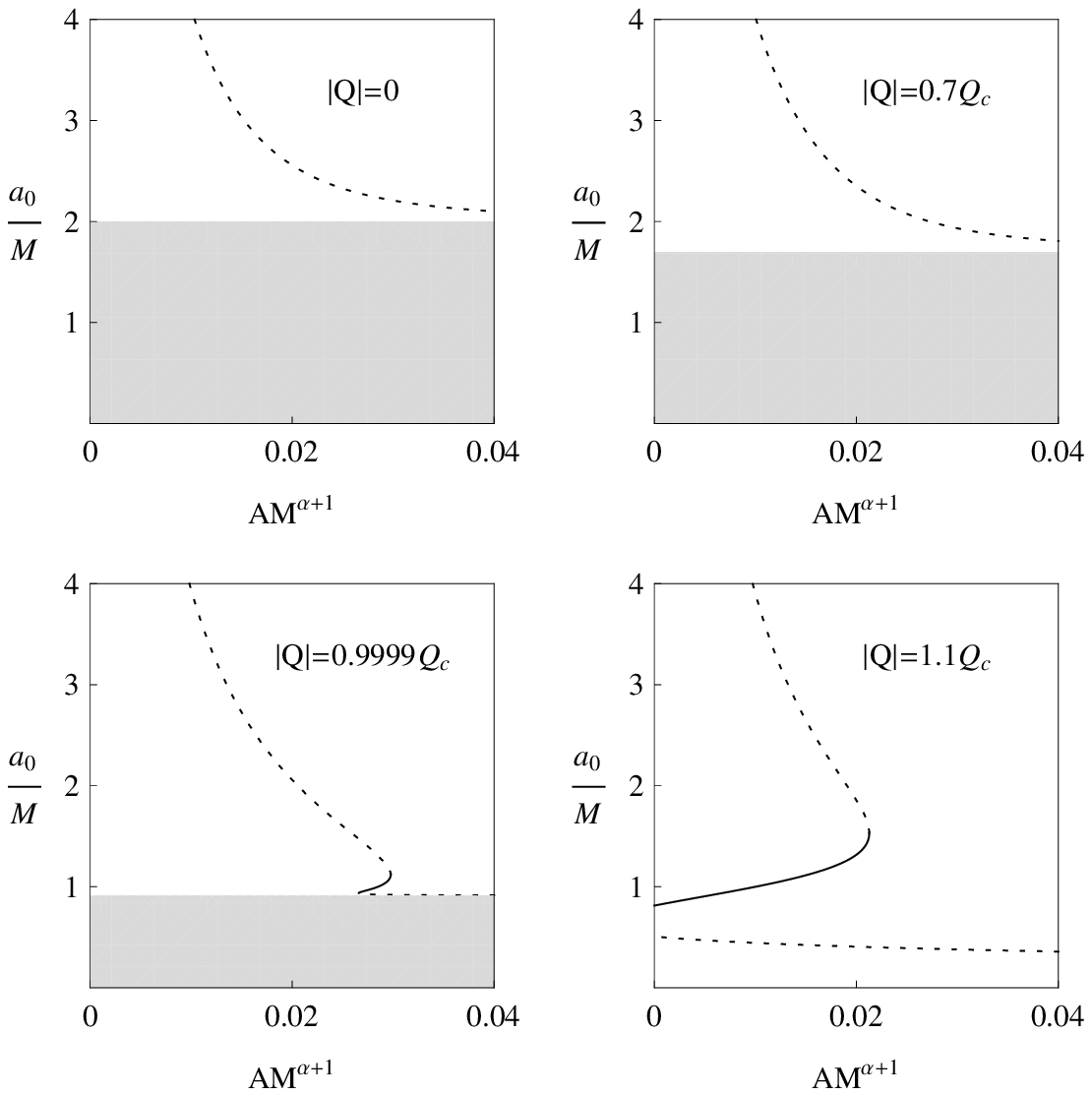}
\caption{Stability of wormholes with parameters $b/M=1$ and $\alpha=0.2$.  In this case, $Q_c/M=1.02526$. Solid (dotted) curves represent static stable (unstable) solutions with throat radius $a_{0}$. Gray zones are non--physical.}
\label{fb1a02}
\end{minipage}
\hfill
\begin{minipage}{0.475\textwidth}
\centering
\vspace{-1.5 cm}
\includegraphics[width=\textwidth]{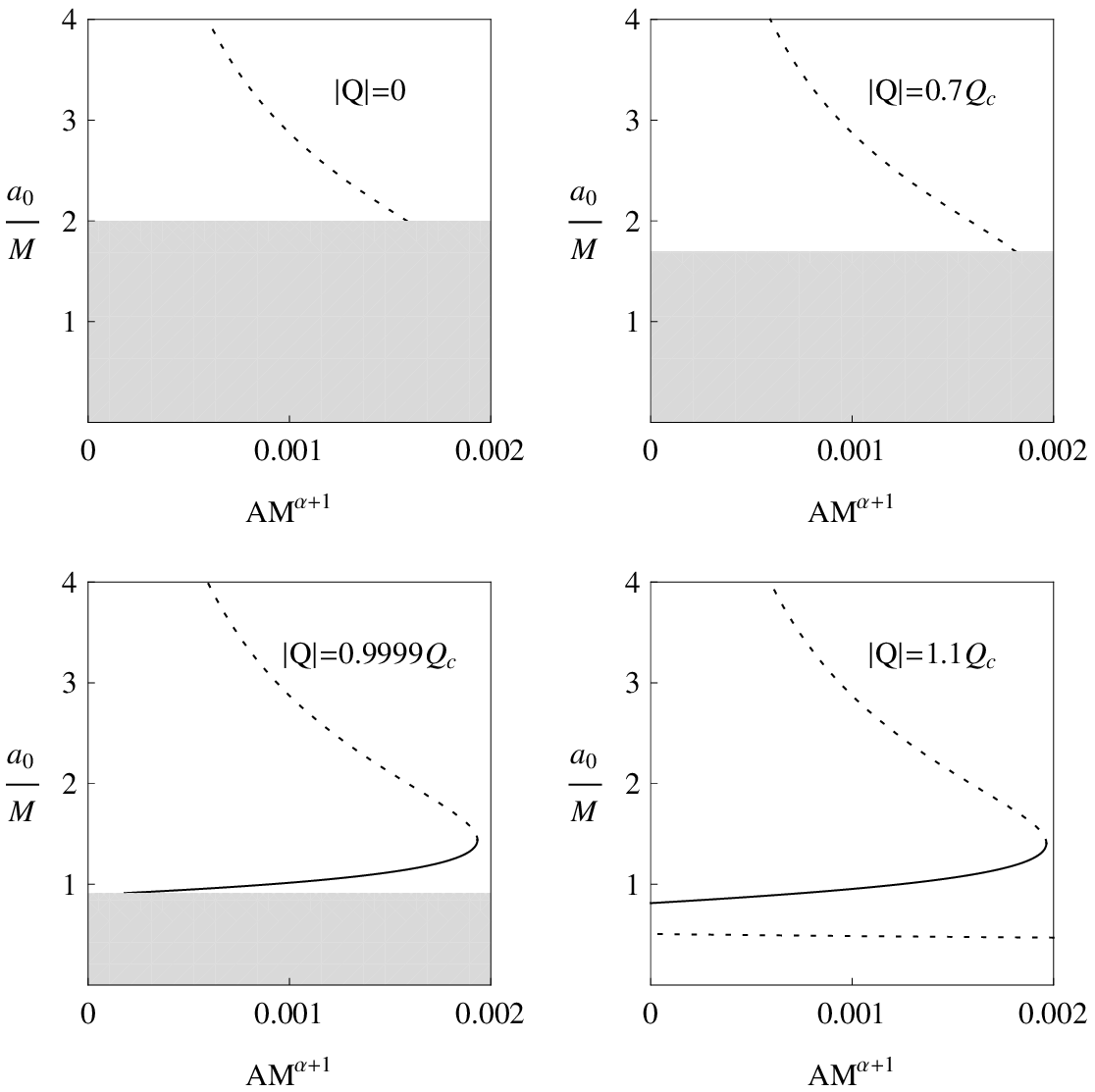}
\caption{Same as Fig. \ref{fb1a02} with $b/M=1$ and $\alpha=1$. In this case, $Q_c/M=1.02526$.}
\label{fb1a1}
\end{minipage}
\end{figure}

\begin{figure}[t]
\centering
\begin{minipage}{0.475\textwidth}
\centering
\includegraphics[width=\textwidth]{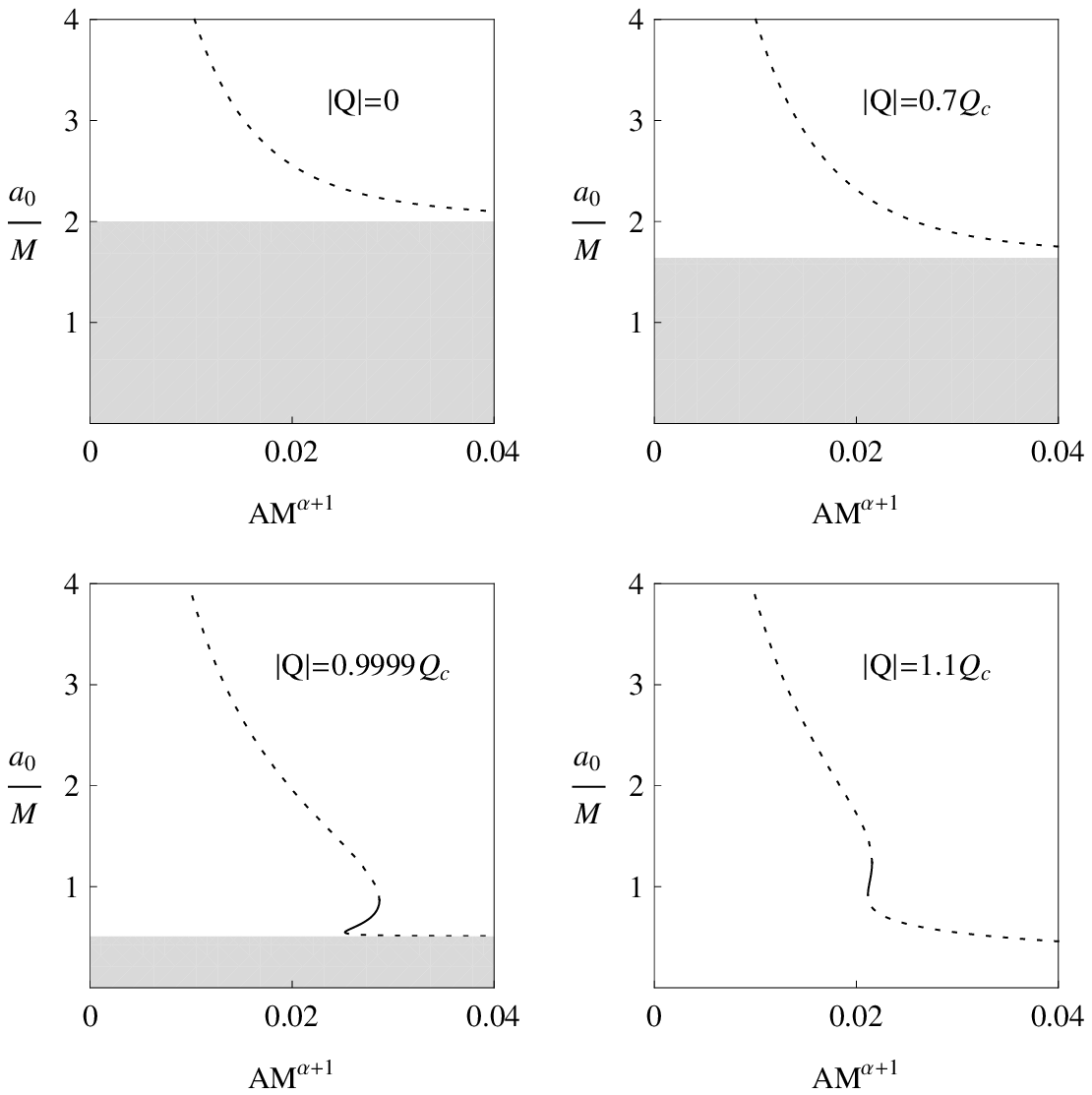}
\caption{Same as Fig. \ref{fb1a02} with $b/M=2$ and $\alpha=0.2$. In this case, $Q_c/M=1.10592$.}
\label{fb2a02}
\end{minipage}
\hfill
\begin{minipage}{0.475\textwidth}
\centering
\includegraphics[width=\textwidth]{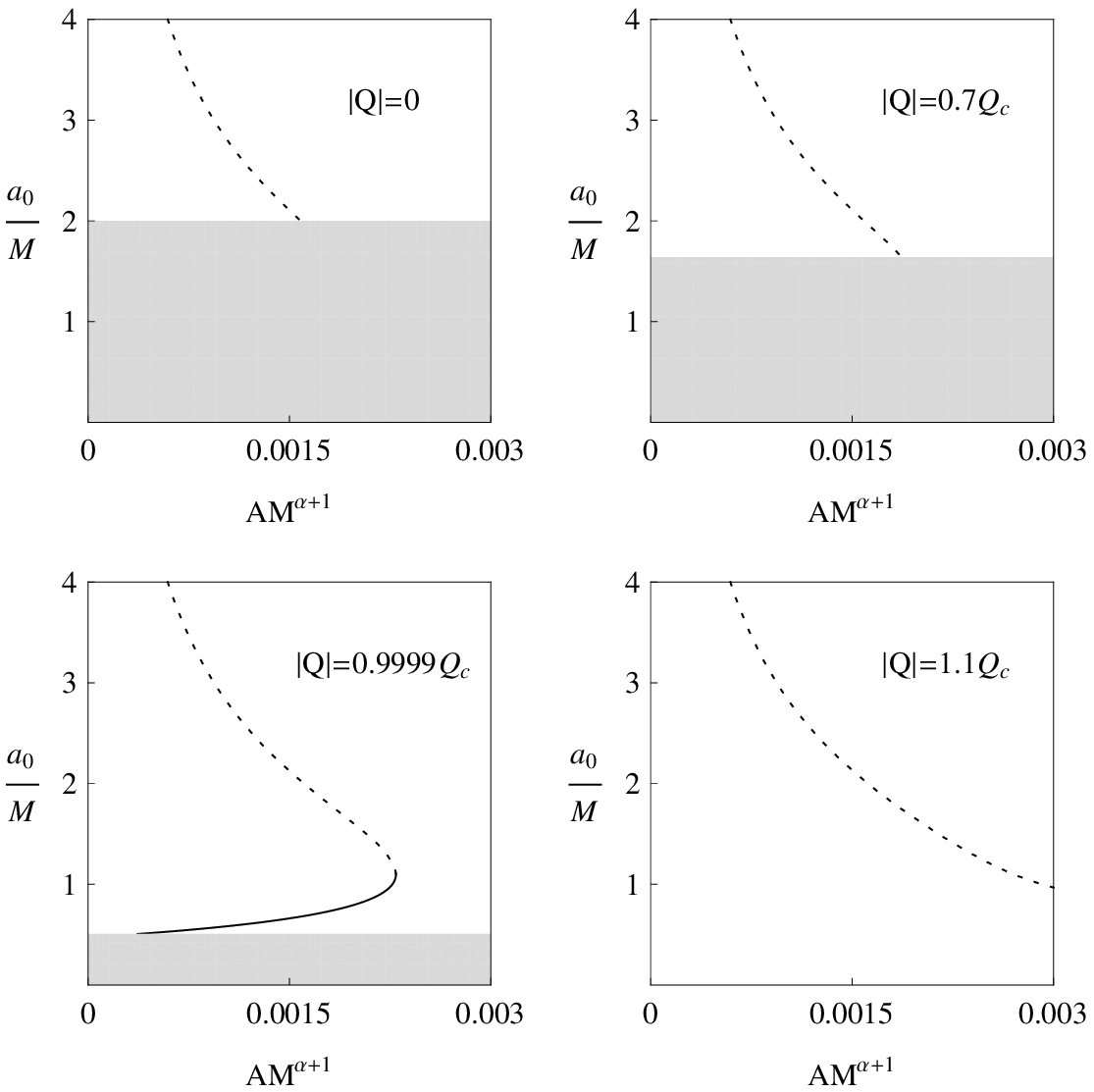}
\caption{Same as Fig. \ref{fb1a02} with $b/M=2$ and $\alpha=1$. In this case, $Q_c/M=1.10592$.}
\label{fb2a1}
\end{minipage}
\end{figure}

\begin{figure}[t]
\centering
\begin{minipage}{0.475\textwidth}
\centering
\includegraphics[width=\textwidth]{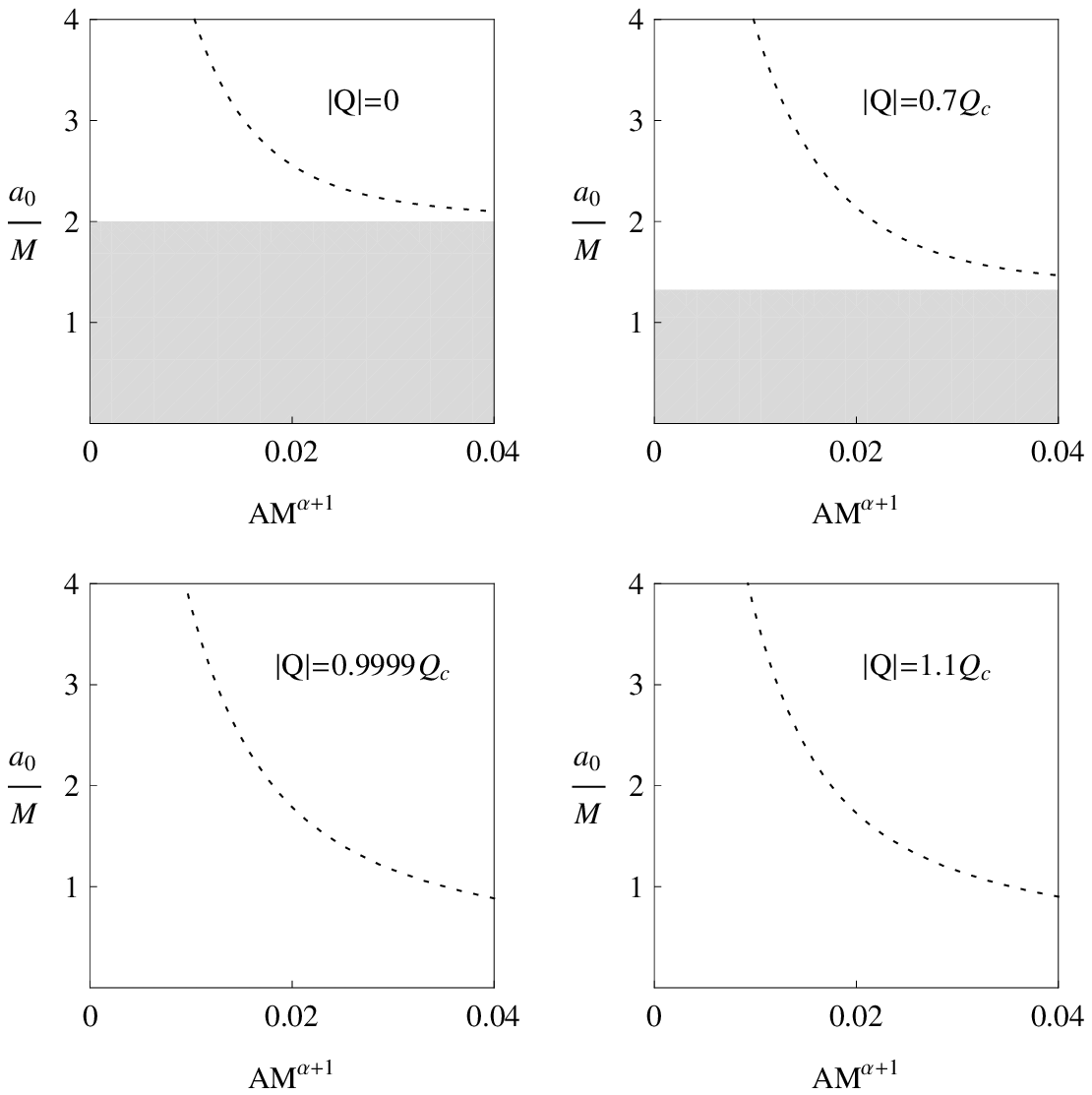}
\caption{Same as Fig. \ref{fb1a02} with  $b/M=5$ and $\alpha=0.2$. In this case, $Q_c/M=1.48468$.}
\label{fb5a02}
\end{minipage}
\hfill
\begin{minipage}{0.475\textwidth}
\centering
\includegraphics[width=\textwidth]{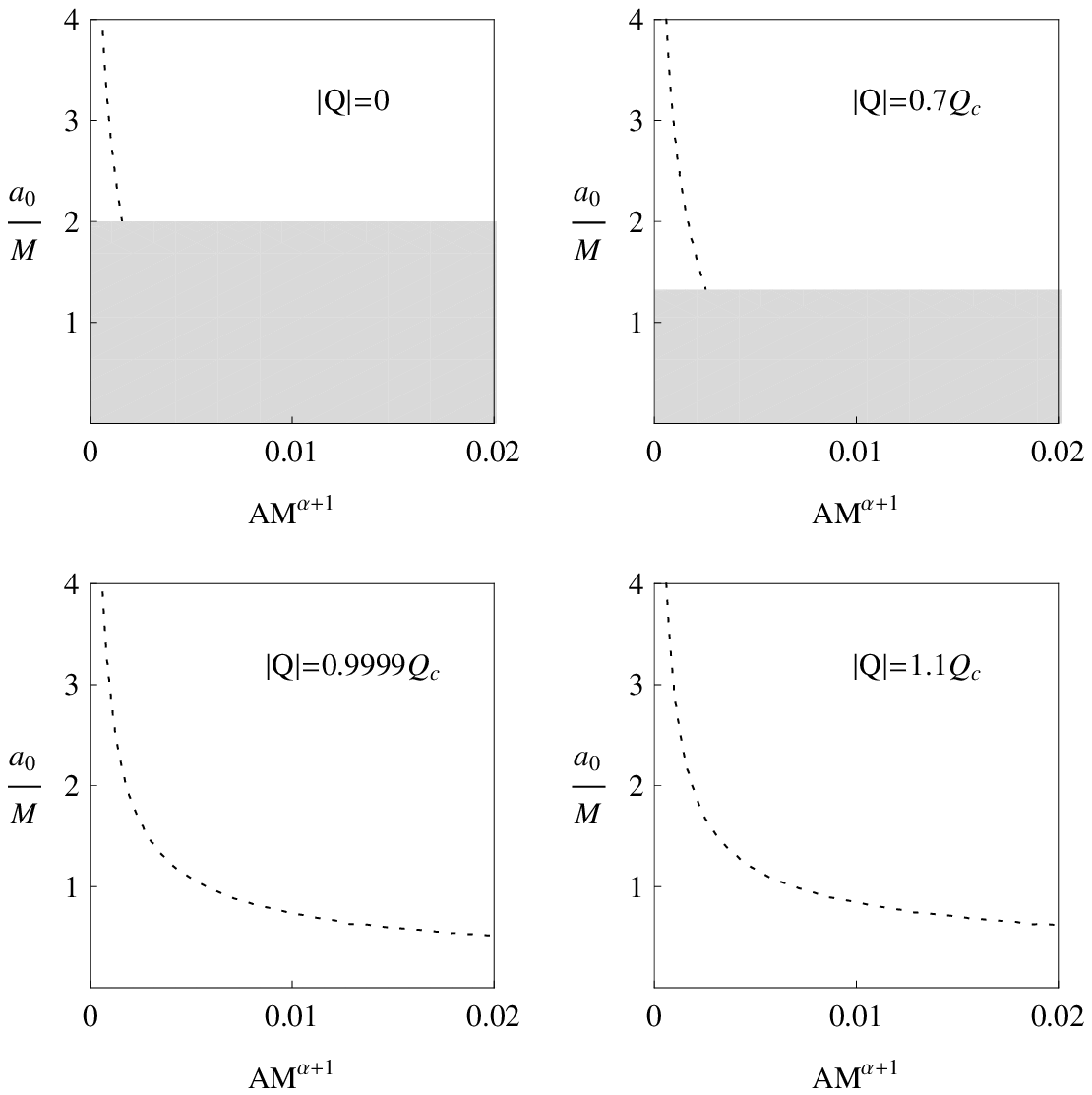}
\caption{Same as Fig. \ref{fb1a02} with $b/M=5$ and $\alpha=1$. In this case, $Q_c/M=1.48468$.}
\label{fb5a1}
\end{minipage}
\end{figure}

From Figs. \ref{fb1a02} to \ref{fb5a1}, we see that:
\begin{itemize}
\item  As discussed above, for $b=0$ (then $Q_c/M=1$), the Born--Infeld electrodynamics reduces to Maxwell theory, so that the Reissner--Nordstr\"{o}m solution is used in the construction of the wormholes, and we recover the results obtained in Ref. \cite{chaply09}.
\item  If $0<b/M\leq1$ the behavior of the solutions is similar to what shown in Figs. \ref{fb1a02} and \ref{fb1a1}, for $b/M=1$ (then $Q_c/M=1.02526$):
\begin{itemize}
\item  When $0<\alpha<1$ (for example, $\alpha = 0.2$):
\begin{itemize}
\item  For $0\leq |Q|<Q_c$ and $|Q|$ not very close to $Q_c$, there is one unstable solution for each value of $AM^{\alpha +1}$, and this behavior continues when this parameter grows.
\item  For $|Q|\lesssim Q_c$, a small range in the values of $AM^{\alpha +1}$ is found for which three solutions are obtained: two of them are unstable (the largest and the smallest ones) and the other is stable. When $AM^{\alpha +1}$ grows, the behavior described above is obtained again: there is only one unstable solution close to the radius of the horizon of the original manifold.
\item  For $|Q|>Q_c$, the range of values of $AM^{\alpha +1}$ where there are three solutions becomes larger; two of them are unstable (the largest and the smallest ones) and the other one is stable. For large values of $AM^{\alpha +1}$ there is again only one unstable solution.
\end{itemize}
\item  When $\alpha=1$:
\begin{itemize}
\item  For $0\leq |Q|<Q_c$ and $|Q|$  not very close to  $Q_c$, we see that a bounded range of values of $AM^{\alpha +1}$ exists for which there is only one unstable solution. The value of $a_0/M$ decreases with $AM^{\alpha +1}$ until it reaches the radius of the horizon of the original manifold, and then no solutions are found.
\item  For $|Q|\lesssim Q_c$, there is a range of values of $AM^{\alpha +1}$ for which two solutions are found, one is stable (the smallest) and the other is unstable (the largest). From a certain value of $AM^{\alpha +1}$ solutions are not longer present.
\item  For $|Q|>Q_c$, there are three solutions, two of them unstable (the largest and the smallest ones) and the other stable. For large values of $AM^{\alpha +1}$ there is only one solution, which is unstable.
\end{itemize}
\end{itemize}
Comparing the results shown in Figs. \ref{fb1a02} and \ref{fb1a1} with those obtained in Ref. \cite{chaply09}, we can see a similar behavior in the cases $b=0$ and $0<b/M\leq 1$, for the same values of $\alpha$. The only difference is found when $|Q|>Q_c$: the smallest of the unstable solutions for $0<b/M\leq 1$ is not present if $b=0$.
\item An analogous analysis can be done for Figs. \ref{fb2a02} and \ref{fb2a1}, corresponding to $b/M=2$ (then $Q_c/M=1.10592$):
\begin{itemize}
\item When $0<\alpha<1$ (for example, $\alpha = 0.2$):
\begin{itemize}
\item For $0\leq |Q|<Q_c$ and $|Q|$ not very close to $Q_c$, we observe that from a certain value of $AM^{\alpha +1}$ there is always one unstable solution.
\item For $|Q|\lesssim Q_c$ or $|Q|>Q_c$, there exists a small range of values of $AM^{\alpha +1}$ where three solutions can be found: two of them unstable (the smallest and the largest ones) and the other stable. From a certain value of $AM^{\alpha +1}$ there is only one solution, which is unstable.
\end{itemize}
\item When $\alpha=1$:
\begin{itemize}
\item For $0\leq |Q|<Q_c$ and $|Q|$ not very close to $Q_c$, there is a bounded range of values of $AM^{\alpha +1}$ for which only one unstable solution is present. Again, we see that $a_0/M$ decreases until it reaches the radius of the horizon of the original manifold, and then no solutions are found.
\item For $|Q|\thicksim Q_c$, there are two solutions, one stable (the smallest one) and the other unstable (the largest one) for a bounded range of $AM^{\alpha +1}$, and outside from this range no solutions are present. 
\item For $|Q|>Q_c$, there is always only one solution, which is unstable.
\end{itemize}
\end{itemize}
\item From Figs. \ref{fb5a02} and \ref{fb5a1}, in which $b/M=5$ (then  $Q_c/M=1.48468$), we can say that:
\begin{itemize}
\item When $0<\alpha <1$ (for example, $\alpha = 0.2$) and for any value of the charge $|Q|$ there is only one solution, which is always unstable.
\item When $\alpha =1$:
\begin{itemize}
\item For  $0\leq |Q|<Q_c$ and $|Q|$ not very close to $Q_c$, a small range of values of $AM^{\alpha +1}$ can be found for which only one unstable solution is present, with $a_0/M$ decreasing quickly to reach the horizon radius of the original manifold.
\item For $|Q|\lesssim Q_c$ or $|Q|>Q_c$, there is only one solution, which is always unstable.
\end{itemize}
\end{itemize}
This shows that for large values ​​of $b/M$, the stability region disappears completely.
\end{itemize}

Comparing these results with those obtained in the above mentioned work  \cite{chaply09}, we observe that the results obtained for values ​​of $b/M>1$ differ significantly from those with $b=0$.

\section{Summary}\label{conclu}

In this paper, we have constructed spherically symmetric wormholes by using the thin-shell formalism within the framework of Einstein--Born--Infeld theory, with a Chaplygin gas at the surface of union, where the throat is localized. We have analyzed the stability of the wormholes under perturbations that preserve the symmetry. The study has been done analytically and standard software was used to display the results graphically. We found stable solutions for a given set of parameters, namely the radius of the throat $a_0$, the parameter $b$ of Born--Infeld electrodynamics, the parameters $A$ and $\alpha $ of the equation of state corresponding to the generalized Chaplygin gas, the mass $M$, and the charge $Q$. The results were compared with those obtained in a previous work, where a similar study was conducted using the Reissner--Nordstr\"om metric. For small $b/M$, there are values of the other parameters for which the solutions are stable, results similar to the Reissner--Nordstr\"om case, except that in 
Einstein--Born--Infeld a new unstable solution were found in the vicinity of the origin for large values ​​of $Q/M$. As $b/M$ increases, i.e. that the theory is distancing itself more from Einstein--Maxwell, the stability region becomes smaller. For large values ​​of $b/M$ the stable solutions are not longer present.

\section*{Acknowledgments}

This work has been supported by Universidad de Buenos Aires and CONICET.


\begin{thebibliography}{99}

\bibitem{motho} M.S. Morris and K.S. Thorne, Am. J. Phys. \textbf{56}, 395 (1988).

\bibitem{visser} M. Visser, \textit{Lorentzian Wormholes} (AIP Press, New
York, 1996).

\bibitem{hovis} D. Hochberg and M. Visser, Phys. Rev. D \textbf{56}, 4745
(1997); D. Hochberg and M. Visser, Phys. Rev. Lett. \textbf{81}, 746
(1998); D. Hochberg and M. Visser, Phys. Rev D \textbf{58}, 044021 (1998).

\bibitem{viskardad} M. Visser, S. Kar and N. Dadhich, Phys. Rev. Lett. 
\textbf{90}, 201102 (2003).

\bibitem{lst} O.B. Zaslavskii, Phys. Rev. D \textbf{76}, 044017 (2007).

\bibitem{poisson} E. Poisson and M. Visser, Phys. Rev. D \textbf{52}, 7318
(1995).

\bibitem{ishak-eirom-lobo} M. Ishak and K. Lake, Phys. Rev. D \textbf{65}, 044011 (2002); E.F. Eiroa and G.E. Romero, Gen. Relativ. Gravit. \textbf{36}, 
651 (2004); F.S.N. Lobo and P. Crawford, Class. Quantum Grav. \textbf{21}, 391 (2004). 

\bibitem{eir} E.F. Eiroa, Phys. Rev. D \textbf{78}, 024018 (2008).

\bibitem{lmv} N. Montelongo Garcia, F.S.N. Lobo, and M. Visser, Phys. Rev. D \textbf{86}, 044026 (2012).

\bibitem{dilem} G.A.S. Dias and J.P.S. Lemos, Phys. Rev. D \textbf{82}, 084023 (2010).

\bibitem{plane} J.P.S. Lemos and F.S.N. Lobo, Phys. Rev. D \textbf{78}, 044030 (2008).

\bibitem{cil} E. F. Eiroa and C. Simeone, Phys. Rev. D \textbf{81}, 084022 (2010).

\bibitem{borninf} M. Born and L. Infeld, Proc. Roy. Soc. \textbf{A144}, 425 (1934).

\bibitem{hoffmann} B. Hoffmann, Phys. Rev. \textbf{47}, 877 (1935).

\bibitem{nle} E.S. Fradkin and A.A. Tseytlin, Phys. Lett. B \textbf{163}, 123 (1985); E. Bergshoeff, E. Sezgin, C.N. Pope, and P.K. Townsend, Phys. Lett. B \textbf{188}, 70 (1987); R.R. Metsaev, M.A. Rahmanov, and A.A. Tseytlin, Phys. Lett. B \textbf{193}, 207 (1987); A.A. Tseytlin, Nucl. Phys. B \textbf{501}, 41 (1997); D. Brecher and M.J. Perry, Nucl. Phys. B \textbf{527}, 121 (1998).

\bibitem{gibbons} G.W. Gibbons and D.A. Rasheed, Nucl. Phys. B \textbf{454}, 185 (1995).

\bibitem{nora} N. Bret\'{o}n, Class. Quantum Grav. \textbf{19}, 601 (2002).

\bibitem{shellbi} M. G. Richarte and C. Simeone, Phys. Rev. D \textbf{80}, 104033 (2009); \textbf{81}, 109903(E) (2010); E.F. Eiroa and C. Simeone, Phys. Rev. D  \textbf{83}, 104009 (2011).

\bibitem{matt} V. Sahni and A. A. Starobinsky, Int. J. Mod. Phys. D \textbf{9}, 373 (2000); P. J. Peebles and B. Ratra, Rev. Mod. Phys. \textbf{75}, 559 (2003); T. Padmanabhan, Phys. Rep. \textbf{380}, 235 (2003).

\bibitem{aero} S. Chaplygin, Sci. Mem. Moscow Univ. Math. Phys. \textbf{21}, 1 (1904); H.-S-Tien, J. Aeron. Sci. \textbf{6}, 399 (1939); T. von Karman,  J. Aeron. Sci. \textbf{8}, 337 (1941).

\bibitem{chap} A. Kamenshchik, U. Moschella and V. Pasquier, Phys. Lett. \textbf{B511}, 265 (2001); M.C. Bento, O. Bertolami and A.A. Sen, Phys. Rev. D \textbf{66}, 043507 (2002).
 
\bibitem{lobo73} F. S. N. Lobo, Phys. Rev. D \textbf{73}, 064028 (2006).

\bibitem{chaply07} E.F. Eiroa and C. Simeone, Phys. Rev. D \textbf{76}, 024021 (2007).

\bibitem{chaplywh} F. Rahaman, M. Kalam and K. A. Rahman, Mod. Phys. Lett. A \textbf{23}, 1199 (2008); V. Gorini, U. Moschella, A.Yu. Kamenshchik, V. Pasquier, and A.A. Starobinsky, Phys. Rev. D \textbf{78}, 064064 (2008); S. Chakraborty and T. Bandyopadhyay, Int. J. Mod. Phys. D \textbf{18}, 463 (2009); M. Jamil, M.U. Farooq and M.A. Rashid, Eur. Phys. J. C \textbf{59}, 907 (2009); P.K.F. Kuhfittig, Gen. Relativ. Grav. \textbf{41}, 1485 (2009); V. Gorini, A.Yu. Kamenshchik, U. Moschella, O.F. Piattella, and A.A. Starobinsky, Phys. Rev. D \textbf{80}, 104038 (2009); A. Mokeeva, and V. Popov, arXiv:1205.1542. 

\bibitem{chaply09} E.F. Eiroa, Phys. Rev. D \textbf{80}, 044033 (2009).

\bibitem{chaply11} C. Bejarano and E.F. Eiroa, Phys. Rev. D \textbf{84}, 064043 (2011).

\bibitem{daris} G. Darmois, M\'{e}morial des Sciences Math\'{e}matiques, Fascicule XXV (Gauthier-Villars, Paris, 1927), Chap. V; W. Israel, Nuovo Cimento B \textbf{44}, 1 (1966); \textbf{48}, 463(E) (1967).


\end{thebibliography}
\end{document}